\documentclass{article}

\usepackage[final]{nips_2018}

\usepackage[utf8]{inputenc} 
\usepackage[T1]{fontenc}    
\usepackage{hyperref}       
\usepackage{url}            
\usepackage{booktabs}       
\usepackage{amsfonts}       
\usepackage{nicefrac}       
\usepackage{microtype}      
\usepackage[pdftex]{graphicx}
\usepackage{color}
\usepackage{mathtools}
\usepackage{amsmath}
\usepackage{tabularx}
\DeclarePairedDelimiter\abs{\lvert}{\rvert}%

\title{From Satellite Imagery to Disaster Insights}

\author{
  Jigar Doshi \\
  CrowdAI\\
  \texttt{jigar@crowdai.com} \\
  \And
  Saikat Basu \\
  Facebook \\
  \texttt{saikatbasu@fb.com} \\
  \And
  Guan Pang \\
  Facebook \\
  \texttt{gpang@fb.com} \\
}

\begin{document}

\maketitle

\begin{abstract}
  The use of satellite imagery has become increasingly popular for disaster monitoring and response. After a disaster, it is important to prioritize rescue operations, disaster response and coordinate relief efforts. These have to be carried out in a fast and efficient manner since resources are often limited in disaster affected areas and it's extremely important to identify the areas of maximum damage. However, most of the existing disaster mapping efforts are manual which is time-consuming and often leads to erroneous results. In order to address these issues, we propose a framework for change detection using Convolutional Neural Networks (CNN) on satellite images which can then be thresholded and clustered together into grids to find areas which have been most severely affected by a disaster. We also present a novel metric called Disaster Impact Index (DII) and use it to quantify the impact of two natural disasters - the Hurricane Harvey flood and the Santa Rosa fire. Our framework achieves a top F1 score of 81.2\% on the gridded flood dataset and 83.5\% on the gridded fire dataset.
\end{abstract}

\section{Introduction}

In the field of computer vision, semantic segmentation in satellite images~\cite{doshi2018residual, demir2018deepglobe} has been extensively employed to understand man-made features like roads, buildings, land use and land cover types. However, most of these analyses are still limited to static snapshots of data involving images acquired at a single time instance. In order to determine the area impacted by a disaster, we can extend these approaches to time-series data to detect areas of change.

A simple solution to detect change in time-series data is to directly compare raw RGB values of satellite images. However, due to different season, lighting and noise, the pixel values across time-series data can be quite different even in areas with no disaster impact. Therefore, many research efforts have been explored to improve disaster mapping from satellite images~\cite{voigt2016global, voigt2007satellite}. In \cite{voigt2016global}, the authors highlight the use of MODIS to develop models to detect disasters. Similarly, in \cite{voigt2007satellite}, the authors highlight various satellite data sources and efforts for disaster response. More recent approaches have studied the use of CNNs for disaster detection from satellite images \cite{cao2018deep, fujita2017damage, duarte2018satellite, ignatiev2018targeted}. In \cite{cao2018deep}, \cite{fujita2017damage}, \cite{amit2017disaster} and \cite{duarte2018satellite}, the authors use CNNs to detect damaged buildings by using damaged and non-damaged buildings as two classes. However, these approaches rely on building relatively large training datasets for damaged areas which is expensive and not-scalable.

In this work, we propose to locate areas of maximal disaster damage by using man-made features as reference, and detecting change in these features can enable us to determine areas where to focus the relief efforts. We train models based on Fully-Convolutional Neural Networks to detect roads and buildings from satellite imagery, and generate prediction masks in regions with disaster. By computing relative change between multiple snapshots of data captured before and after a disaster, we can identify areas of maximal damage and prioritize disaster response efforts. As our proposed approach compares the change only in high-level man-made features, it's invariant to season, lighting and noise difference in time-series data. In addition, compared to recent approaches~\cite{cao2018deep, fujita2017damage, amit2017disaster, duarte2018satellite} that require training CNNs specifically to detect damaged features, our approach only uses models trained on general road and building datasets, which are relatively inexpensive and hence scalable to other similar natural disasters. Evaluated on a human annotated dataset and a state-provided dataset of actual disaster impacted areas, we show a strong positive correlation between predicted disaster areas and actual disaster impacted areas.

\section{Proposed Approach}

\subsection{Overview}
\begin{figure*}[ht]
\begin{center}
\includegraphics[width=0.99\linewidth]{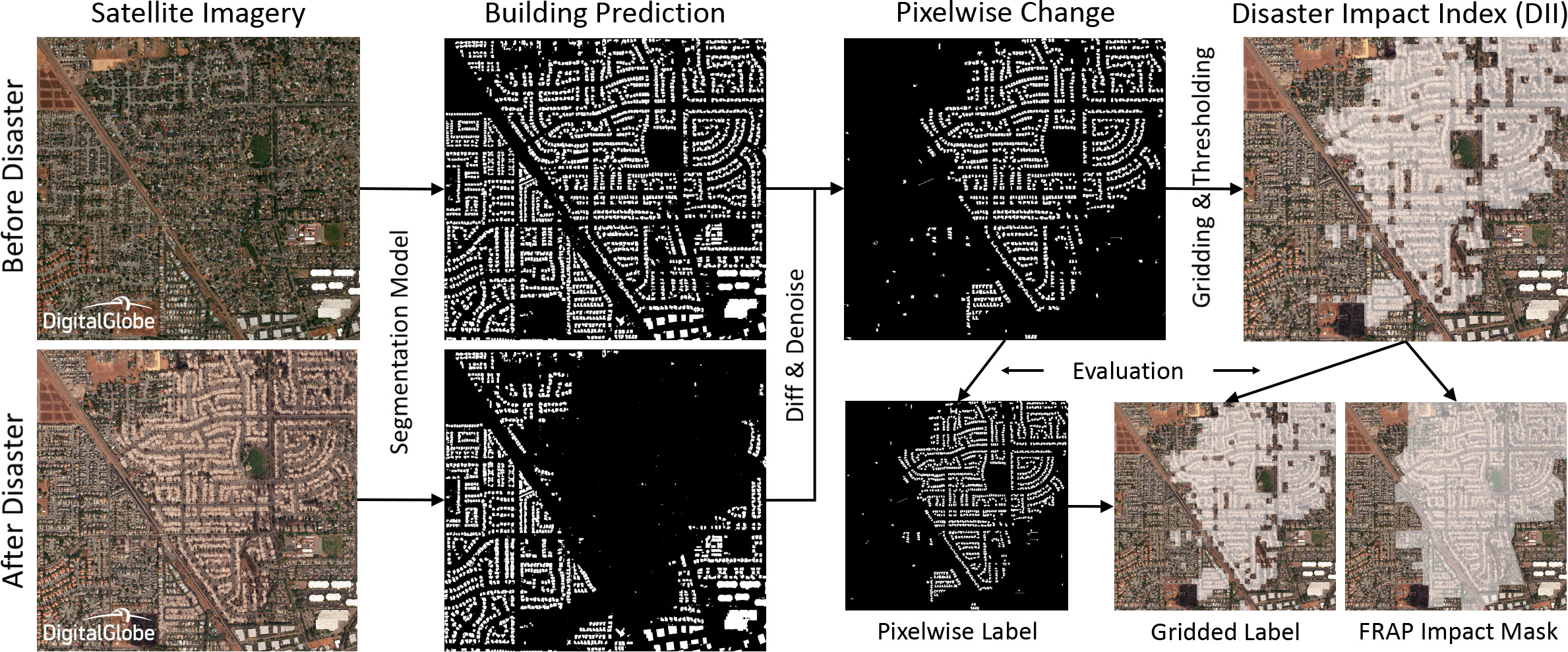}
\end{center}
   \caption{Flow diagram of our proposed approach for disaster impact analysis. We run pre-trained CNN on satellite imagery before and after disaster, compare the change in extracted man-made feature, then compute Disaster Impact Index (DII) to understand impact of each area}
\label{fig:flow}
\end{figure*}

We propose to identify disaster-impacted areas by comparing the change in man-made features extracted from satellite imagery. Using a pre-trained semantic segmentation model (refer to sec.~\ref{sec:model} for details) we extract man-made features (e.g. roads, buildings) on the before and after imagery of the disaster affected area. Then, we compute the difference of the two segmentation masks to identify change. Since the segmentation mask is noisy, we also apply dilation at a radius of $5$ pixels on pre-disaster mask. Subsequently, the change mask is further de-noised by removing small connected components less than $1000$ pixels. Finally, the pixel-wise change mask will be used to compute the Disaster Impact Index (DII - refer to sec.~\ref{sec:metric}) on grids each of size $n{\times}n$ (representing different areas). Figure~\ref{fig:flow} shows a flow diagram of our proposed disaster analysis approach.

\subsection{Model Architecture}
\label{sec:model}
For semantic segmentation model, we use a Residual Inception Skip network following \cite{doshi2018residual}. This model is a convolutional encoder-decoder architecture with the inception modules instead of standard convolution blocks. 
The inception models were originally proposed in \cite{ioffe2015batch} but with asymmetric convolutions. For example, a $3\times3$ convolution is replaced with a $3\times1$ convolution, then batch norm followed by $1\times3$ convolution. This is useful since it reduces the number of parameters and gives similar performance. All weights are initialized with the He norm \cite{he2015delving} and all convolutional layers are followed by batch normalization layers which in turn are followed by activation layers. Following the architecture proposed in \cite{doshi2018residual}, we also used leaky ReLUs with a slope of $-0.1x$ as our activation function. We used a continuous version of the Dice score as our loss function.

\subsection{Disaster Impact Index (DII)}
\label{sec:metric}
As our CNN model detects man-made features before disaster but fails to detect some of them after disaster, we can infer areas of maximal impact using change detection. We propose a metric to quantify this impact called Disaster Impact Index (DII). Considering the semantic segmentation result generated by the CNN before and after the disaster, and dividing the area into small grids each of size $n{\times}n$, we can calculate the change in detected features and define DII in each grid as:

\begin{equation}
    DII = \Delta Pred = \dfrac{{\abs{{\eta}_{Pred_{before} = 1 \& Pred_{after} = 0}}}_{grid}}{\frac{1}{N_{grid}}\sum_{i=1}^{N_{grid}}{\abs{{\eta}_{Pred_{before} = 1}}}_{grid_i}}
\end{equation}

where ${\abs{{\eta}_{Pred_{before} = 1 \& Pred_{after} = 0}}}_{grid}$ denotes the number of pixels in the grid which have the feature detected in the pre-disaster CNN mask but not in the post-disaster mask (because the other way around is usually not caused by disaster); $\frac{1}{N_{grid}}\sum_{i=1}^{N_{grid}}{\abs{{\eta}_{Pred_{before} = 1}}}_{grid_i}$ denotes the number of feature pixels predicted in each grid pre-disaster, averaged over the whole region, where $N_{grid}$ is the total number of grids in the region. Essentially, DII represents the normalized pixelwise change aggregated over the evaluation grid. This normalization is crucial since the number of feature pixels varies depending on the region (urban, rural) or feature (road, building). After normalization, DII becomes a region and feature independent metric that's comparable across different scenarios.

Aggregating the DII over grids of arbitrary sizes, and thresholding with a threshold of $\tau$, we can infer regions of maximal impact. In our analysis, we set grid size to be $256\times256$, and $\tau=0.01$. Note the threshold is the same regardless of region or feature, thanks to the normalization in DII.

\section{Experiments}

\subsection{Training Datasets}

We trained our model by combining two publicly available high-resolution satellite imagery semantic segmentation datasets, namely Spacenet~\cite{spacenet} and Deepglobe~\cite{demir2018deepglobe}. Spacenet is a corpus of commercial satellite imagery and labeled training data which consists of building footprints for various cities around the world at resolutions ranging from 30-50 cm/pixel. The DeepGlobe dataset is created from DigitalGlobe Vivid+ satellite imagery~\cite{digitalglobe} containing roads, buildings and landcover labels at resolution of 50 cm/pixel. To show that our method generalizes across feature types and datasets, we also used another dataset of lower resolution imagery (around 3 m/pixel) from Planet Labs~\cite{planet} to train the roads model.

\subsection{Validation Datasets}

In order to validate our results we identified two natural disasters: Hurricane Harvey flood \cite{planetharvey} and Santa Rosa fire \cite{dgsantarosa}. The Hurricane Harvey flood dataset is approximately $143km^2$ near Sugar Land, Texas; and Santa Rosa fire dataset is approximately $120km^2$ near Santa Rosa, California. These areas were chosen to represent two classes of natural disasters, namely flood and fire whose impact is evaluated using two classes of man-made structures, namely roads and buildings respectively. They also represent two different resolution datasets, that is DigitalGlobe and Planet Labs.  

We annotated the data in two different ways. Firstly, we followed the same annotation procedure as described in the DeepGlobe paper \cite{demir2018deepglobe} to identify all the roads and buildings in a pixel-wise binary mask. Secondly, instead of completely relying on pixel based metrics which may not fully capture the goal of impact area analysis, we split the image in small grids and asked the annotators to identify affected area from satellite imagery . For Santa Rosa, we found ground truth data from the FRAP website from the California Department of Forestry and Fire Protection \cite{frap}, so we used that instead of the human annotated version. 


\begin{figure*}[ht]
\begin{center}
\includegraphics[width=0.99\linewidth]{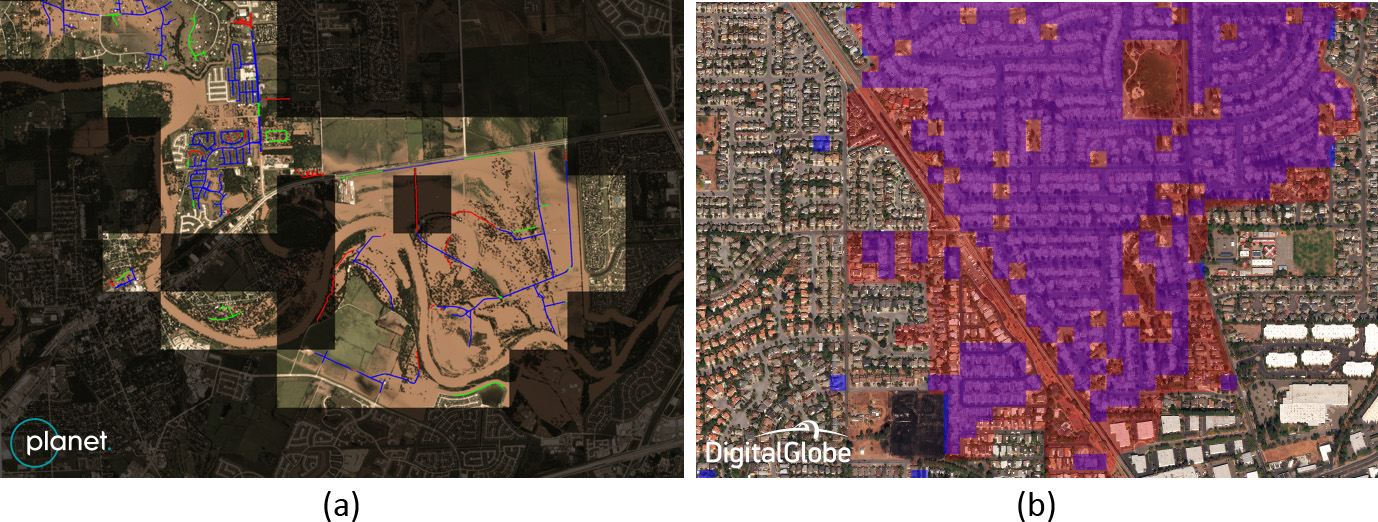}
\end{center}
   \caption{(a) Use DII to infer severe flooding damage areas (highlighted) during Hurricane Harvey, based on changes in detected roads (Blue: true positive, Green: false positive, Red: false negative); (b) Use DII to infer severe fire damage areas at Santa Rosa, based on changes in detected buildings (Blue), compared with shapefile of fire damage area obtained from FRAP \cite{frap} (Red)}
\label{fig:analysis}
\end{figure*}

\subsection{Evaluation Results}
\label{results}

For both the Hurricane Harvey flood data as well as the Santa Rosa fire data, we evaluate the proposed approach in three settings as shown in fig.~\ref{fig:flow}: i) predict pixelwise change before and after disaster, and evaluate against pixelwise ground-truth labels, ii) aggregate pixelwise changes over grids of size $256\times256$ to compute DII, and threshold by $\tau=0.01$, then evaluate against gridded ground-truth computed from pixelwise labels, and 3) compute DII and threshold the same way, but evaluate against labels for impact area.

Figure~\ref{fig:analysis}(a) shows a flood affected area impacted during Hurricane Harvey. We can see several roads missing after the flood, which can help us quantify impact of the flood. Finding areas of maximal and minimal change on top of these disaster affected areas, and thresholding and clustering them, we can infer areas of most severe flooding as highlighted in fig.~\ref{fig:analysis}(a). The quantitative results from this analysis are presented in Table~\ref{table:harvey}. Figure~\ref{fig:analysis}(b) shows change in building structures detected before and after the Santa Rosa fire, and quantitative results are presented in Table~\ref{table:fire}. Using the human annotated dataset of actual disaster impacted areas for the Harvey flood and the FRAP dataset for the Santa Rosa fire, we are able to prove a positive correlation between DII and actual disaster impacted areas. This can be seen in Table~\ref{table:harvey} and ~\ref{table:fire} where the F1 scores indicate the high correlation between the CNN-based change detection masks and the ground truth data both for the pixelwise approach and the DII-based approach.

\begin{table}[ht]
\scriptsize
\renewcommand{\arraystretch}{1.25}
\centering
    \caption{Results for the evaluation area for the Hurricane Harvey flood}
    \begin{tabularx}{\textwidth}{lcccc}
        \hline
        & Precision & Recall & F1  & IoU  \\
        & & & \cite{sasaki2007truth} & \cite{rahman2016optimizing} \\
        \hline
        Pixelwise road change prediction vs. pixelwise labelled roads & 63.1\% & 67.9\% & 65.4\% & 48.6\% \\
        \hline
        DII-based road change prediction vs. gridded labelled roads & 75.9\% & 87.2\% & 81.2\% & 68.3\% \\
        \hline
        DII-based road change prediction vs. labelled impact area & 88.8\% & 50.5\% & 64.4\% & 47.5\% \\
        \hline
    \end{tabularx}
    \label{table:harvey}
\end{table}

\begin{table}[ht]
\scriptsize
\renewcommand{\arraystretch}{1.25}
\centering
    \caption{Results for the evaluation area for the Santa Rosa fire}
    \begin{tabular}{lcccc}
        \hline
        & Precision & Recall & F1 & IoU \\
        \hline
        Pixelwise building change prediction vs. pixelwise labelled buildings & 72.4\% & 81.7\% & 76.8\% & 62.3\% \\
        \hline
        DII-based building change prediction vs. gridded labelled buildings & 81.8\% & 85.4\% & 83.5\% & 71.7\% \\
        \hline
        DII-based building change prediction vs. FRAP impact area & 81.1\% & 73.5\% & 77.1\% & 62.7\% \\
        \hline
    \end{tabular}
    \label{table:fire}
\end{table}

\section{Conclusion and Future Work}

In this paper, we show that using multiple snapshots of satellite images captured at different time periods, and running CNN-based semantic segmentation models, we can detect change in the structure of various man-made features and use this as a proxy to detect areas of maximal impact due to natural disasters. The change masks derived from CNN outputs are clustered and thresholded to derive the Disaster Impact Index (DII) which can be used to find regions of priority to coordinate relief efforts. Our experiments show a high correlation between predicted impact areas and ground truth labels.


As part of this work, we focus only on roads and buildings, however this can be extended to quantify disaster impact on other general natural and man-made features.

\section*{Acknowledgement}
We would like to thank the CrowdAI team especially Pablo Garcia for providing us with the labels and to DigitalGlobe and Planet Labs for making their satellite imagery available.   

\bibliographystyle{authordate1}
\bibliography{nips_2018}

\end{document}